\documentclass[epj]{svjournew}
\usepackage{graphics}
\newcommand{\lsim}{
\mathrel{\hbox{\rlap{\hbox{\lower4pt\hbox{$\sim$}}}\hbox{$<$}}}}
\begin{document}
\title{CKM angles from non-leptonic B decays\\ using SU(3) flavour 
symmetry}
\author{Joaquim Matias
}                     
%
%
\institute{Universitat Autonoma Barcelona, 08193 Bellaterra, Barcelona, 
Spain\\}
%
%
\abstract{
We discuss the determination of the CKM angles $\gamma$ and $\alpha$ 
using recent data from non-leptonic B decays together with flavour 
symmetries. Penguin effects are controlled by means of the CP-averaged 
branching ratio $B_d \to \pi^\pm K^\mp$. The information from ${\cal 
A}_{CP}(B_d \to J/\Psi K_S)$ (two solutions for $\phi_d$), $R_b$ and 
$\gamma$ allow us to determine 
$\beta$, even in presence of New Physics not affecting $\Delta B=1$ 
amplitudes. In this context we  address 
the question of to what extent there is still space for New Physics.
\PACS{
      {13.25Hw}{Hadronic decays of mesons}   \and
      {11.30Er}{CP violation}
     } 
} 
\maketitle
\section{Introduction}
\label{intro}
B physics is  one of the most fertile testing grounds to check the CKM 
mechanism of CP violation in the SM~\cite{revs}, but also to look for the 
first 
signals of New Physics~\cite{new} in the pre-LHC era. 

The huge effort at the 
experimental level at the B factories and future hadronic 
machines~\cite{hadronic}
has produced, already, several impressive results. First, the measurement 
of ${\sin 
\phi_d}$ from the mixing induced CP asymmetry of the decay $B_d \to J/\Psi 
K_S$. 
Second, the measurement of a 
series of non-leptonic B decays: $B_d \to \pi K$, $B_d \to \pi \pi$ and in 
the 
future hadronic machines $B_s \to KK$ will be also accessible.

These non-leptonic B decays play a fundamental role in the determination 
of the CKM angle $\gamma$. The main problem in analysing them
is how to deal with hadronic matrix elements and how to control penguin 
contributions. Our approach~\cite{fl1,FlMa1,FlMa2,RF-Bpipi} extract the 
maximal 
possible information from 
data using flavour symmetries to try to reduce as much as possible the 
uncertainties associated 
to QCD hypothesis.

\section{CKM angle $\gamma$ from non-leptonic decays: $B_d\to \pi \pi $, 
$B_d \to \pi 
K$ and
$B_s \to KK$}
\label{sec:1}
We start writing down a general amplitude parametrization 
of $B_d \to \pi^+ \pi^-$ in the SM~\cite{fl1,FlMa2}:
$$\begin{array}{ll}
A(B_d^0 \rightarrow \pi^+
\pi^-)={\cal 
C}\left(e^{i 
\gamma} -  {\bf d
e^{i\theta}}\right) 
\end{array}$$
All the hadronic information is collected in 
$${ d e^{i\theta}}\equiv\frac{1}{R_b}\left(
\frac{{ A_{\rm pen}^{ct}}}{{
A_{\rm
CC}^{u}}+{ A_{\rm pen}^{ut}}}\right)
\quad
{\cal C}\equiv\lambda^3A\,R_b\left({ A_{\rm
CC}^{u}}+{ A_{\rm
pen}^{ut}}\right)$$
where $A_{\rm
CC}^{u}$ are current-current contributions and  $A_{\rm
pen}^{qt}$ are differences between penguin contributions with a quark 
$q=u,c$ 
and 
a 
quark top inside the loop.

This amplitude allow us to construct the corresponding CP 
asymmetries~\cite{fl1,FlMa2}:
$$
{{\cal A}_{\rm
CP}^{{\rm dir}}}= {\rm func}({ d},
{\theta}, { \gamma}) \quad  {{\cal A}_{\rm
CP}^{{\rm mix}}}={\rm func}({ d},
{\theta}, { \gamma}, { \phi_d})
$$
Following a similar procedure we can write down the amplitude for a 
closely 
related process:
$$
{A(B_s^0
\rightarrow K^+ K^-)=}
\left(\frac{\lambda}{1-\lambda^2/2}\right) {\cal 
C}^\prime\left[e^{i 
\gamma}+\left(\frac{1-\lambda^2}{\lambda^2}\right) 
{ d^\prime} e^{i\theta'}\right]
$$
whose corresponding asymmetries will depend on~\cite{fl1,FlMa2}:
$$
{ {\cal A}_{\rm CP}^{{\rm dir}}}=
{\rm func}({ d^\prime},\theta^\prime,{\gamma}) \quad
{\cal A}_{\rm CP}^{{\rm mix}}=
{\rm func}({ d^\prime},\theta^\prime,\gamma,{
\phi_s})
$$
The crucial point, here, is that the hadronic parameters $d^\prime$, 
$\theta^\prime$ and ${\cal 
C}^\prime$, has exactly the same functional dependence on the penguins 
that $d$,
$\theta$ and ${\cal
C}$, 
except for the interchange of a {\it d} quark by an {\it s} quark.

As a consequence, both processes can be related via U-spin symmetry, 
reducing the total number of parameters to five: $\gamma$, $d$, $\theta$, 
$\phi_d$ and $\phi_s$. At this point, one must 
check the sensitivity of the results to the breaking of  U-spin symmetry. 
This is 
explained in subsection~\ref{sec:22}.  

Looking a bit more in detail, one finds that $d$ is indeed not a free 
parameter, but it can be constrained or substituted using an observable 
called $H$~\cite{RF-Bpipi,FlMa2}:
$$
H\!\equiv\!\frac{1}{\epsilon}\!\left|\frac{{\cal C}'}{{\cal
C}}\right|^2\!
\left[\frac{M_{B_d}}{M_{B_s}}\,\frac{\Phi(\frac{M_K}{M_{B_s}},
\frac{M_K}{M_{B_s}})}{
\Phi(\frac{M_\pi}{M_{B_d}},\frac{M_\pi}{M_{B_d}})}\,\frac{\tau_{B_s}}{\tau_{B_d}}\right]
\!\left[\frac{\mbox{BR}(B_d\to\pi^+\pi^-)}{\mbox{BR}(B_s\to
K^+K^-)}\right]
$$
This quantity requires the knowledge of ${\rm BR}(B_s\to
K^+K^-)$, which is still not available. However, we can already now 
evaluate $H$ by 
making 
contact with the B factories and substitute $B_s \to K^+K^-$ by $B_d 
\to \pi^\pm K^\mp$. These two processes differ by the spectator quark and 
certain exchange and penguin annihilation topologies that are expected to 
be small~\cite{ghlr}. This leads to the following value for 
$H$~\cite{FIM}:
\begin{equation} \label{hexp}
H\approx\frac{1}{\epsilon}\left(\frac{f_K}{f_\pi}\right)^2
\left[\frac{\mbox{BR}(B_d\to\pi^+\pi^-)}{\mbox{BR}(B_d\to\pi^\mp
K^\pm)} \right]=7.5 \pm 0.9
\end{equation}
Due to the dependence of $H$ only on ${\rm cos}\theta\, {\rm cos} \gamma$
in the U-spin limit, we obtain immediately a constrained range for 
$d$: 
$0.2\leq d \leq 
1$. Also, using the exact expression for $H$ we can obtain $d$ as a 
function of $H$, 
$\theta$ and $\gamma$.

It is important to insist here that once the data on the branching ratio 
of $B_s \to KK$ will be available, the spectator quark hypothesis will not 
be necessary and only U-spin breaking effects will be important.

\subsection{Prediction for CKM-angle $\gamma$}
\label{sec:21}

Let's take as starting point the general expression~\cite{FlMa2}:
\begin{equation}\label{start}
\!{{\cal A}_{\rm CP}^{\rm
dir}(B_d\!\to\!\pi^+\!\pi^-)\!=\!}\!\mp\!\left[
\frac{\sqrt{4d^2-\left(u+vd^2\right)^2}\sin{\gamma}}{(1-
u\cos{\gamma})+
(1-v\cos{\gamma})d^2}\right]\!
\end{equation}
where
$u,v,d=F_i( {\cal A}_{\rm CP}^{\rm mix}, H,
 \gamma,  \phi_d (B_d \to J/\Psi
K_s);  \xi, \Delta \theta)$. The parameters $\xi$, $\Delta \theta$
will account for the U-spin breaking and are discussed in 
subsection~\ref{sec:22}.

Using present world average for ${\rm sin}\phi_d=0.734\pm0.054$, one 
obtains 
two possible
solutions for the weak mixing angle:
$$\label{phid-det}
\phi_d=\left(47^{+5}_{-4}\right)^\circ \, \lor \,
\left(133^{+4}_{-5}\right)^\circ.
$$
We will refer later on to these two solutions like  scenario A 
and B, 
respectively.

Concerning experimental data, the situation is still uncertain, but 
improving. Present 
naive average of Belle and Babar data is~\cite{expe}:
\begin{eqnarray}
{{\cal A}_{\rm CP}^{\rm dir}(B_d\to\pi^+\pi^-)}&=&-0.38\pm0.16
\,\,  \nonumber
\label{Bpipi-CP-averages}\\
{{\cal A}_{\rm CP}^{\rm mix}(B_d\to\pi^+\pi^-)}&=&+0.58\pm0.20
\,\,  \nonumber
\label{Bpipi-CP-averages2}
\end{eqnarray}
The intersection of the two experimental ranges of ${\cal A}_{\rm CP}^{\rm 
dir}$ and ${\cal A}_{\rm CP}^{\rm mix}$ allow us, using Eq.~(\ref{start}), 
to 
determine the 
range
for 
$\gamma$. The first range, corresponding to take $\phi_d=47^\circ$ is:
\begin{equation} \label{range1}
32^\circ \lsim \gamma \lsim 75^\circ \end{equation}
For the second solution $\phi_d=133^\circ$ one obtains:
\begin{equation} \label{range2}
105^\circ \lsim \gamma \lsim 148^\circ \end{equation}    
Both plots are symmetric (see~\cite{FlMa2,meu}). This is a consequence of 
the symmetry 
$\phi_d \rightarrow 180^\circ
-\phi_d$, $\gamma \rightarrow 180^\circ - \gamma$
that Eq.~(\ref{start}) exhibits.
It is remarkable the stability of the range for $\gamma$ if we compared 
it with 
previous analysis~\cite{meu}.

\subsection{Sensitivity to parameters $H$, $\xi$ and $\Delta \theta$}
\label{sec:22}

Here we will analyze the sensitivity of the determination of $\gamma$ on 
the 
variation of the different hadronic parameters.

\subsubsection{$H$ and the spectator quark hypothesis}

Let's fix the solution $\phi_d=47^\circ$ and take the 
experimental branching ratios of  
$B_d \to \pi \pi$ and $B_d \to \pi K$ to determine $H$. We vary $H$ 
inside 
its 
experimental range Eq.~(\ref{hexp}) at one, two and three sigmas to take 
into 
account the 
uncertainty associated to the spectator quark hypothesis. We find at one 
sigma a very mild influence in the determination of 
$\gamma$. The error 
induced in the range of $\gamma$ is about $\pm 2^\circ$. 

For the very conservative range of up to three sigmas we find a maximal 
error 
of $6^\circ$. Moreover, if the experimental value of $H$ tends to increase 
the range for $\gamma$ tends to decrease, allowing for a narrower
determination.

Finally, the uncertainty associated to $H$ will be drastically reduced 
once 
the ${\rm BR}(B_s \to KK)$ is known and $H$ will be taken safely in a 
narrower range.

\subsubsection{U-spin breaking: $\xi$ and $\Delta \theta$}      

U-spin breaking is the most important uncertainty. We will follow two 
different strategies to keep it under control:

\begin{itemize}
\item[a)] Once the data from the CP asymmetries and branching ratio of 
$B_s \to KK$ will be 
available and $\phi_s$ will be measured from the CP-asymmetry of $B_s \to 
J/\Psi \phi$, we 
will be able to {\it test} directly from data U-spin breaking. Taking 
$\phi_d$ from $B_d \to J/\Psi K_S$ we will have 4 observables (the CP 
asymmetries) and 3 unknowns ($d$, $\theta$, $\gamma$). Then, we can add 
$d^\prime$ as another free parameter and data will tell us the amount of 
U-spin breaking.
\medskip

\item[b)] Already now, we can define two quantities $\xi=d^\prime/d$ and 
$\Delta \theta=\theta^\prime-\theta$ that parametrizes the amount of 
U-spin 
breaking. In order to test the sensitivity of $\gamma$ 
to the variation of these parameters, we  allow them to vary in a 
range. If we allow for a very large variation of $\xi$ between 
$0.8$ 
and $1.2$, the larger error in the determination of $\gamma$ is of $\pm 
5^\circ$. Concerning $\Delta \theta$, its influence is negligibly small, 
a variation of $40^\circ$ induces an error of at most 1 degree.

\end{itemize}

Other studies on U-spin breaking can be found in~\cite{beneke}.

\begin{figure}
\resizebox{0.48\textwidth}{!}{%
  \includegraphics{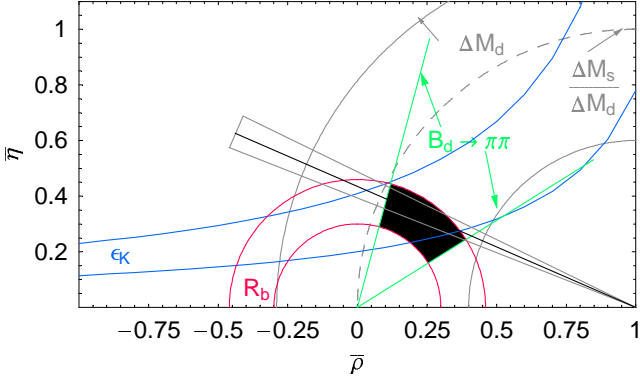}
}
\caption{$\phi_d=47^\circ$. SCENARIO A}
\label{fig:1}       
\end{figure}



\section{Determination of CKM angles $\alpha$ and $\beta$ in SM
and with New Physics in the mixing}

Next point is how to
determine $\alpha$ and $\beta$~\cite{FIM}.
Here, in addition,  we will also allow  for Generic New Physics affecting
the
$B^0_d$--$\overline{B^0_d}$ mixing, but not to the $\Delta
(B,S)=1$
decay amplitudes, i.e, this type of New Physics is consistent with the 
determination of $\gamma$ explained in the previous section. Our inputs 
are\cite{FIM,GNW}: \begin{itemize}
\item ${R_b}\equiv\left|\frac{V_{ud}V_{ub}^\ast}{
V_{cd}V_{cb}^\ast}\right|
$ obtained from exclusive/inclusive transitions mediated by  $b\to u
\ell\overline{\nu}_\ell$ and
$b\to c\ell\overline{\nu}_\ell$. Two important remarks are:
a) This is an observable practically insensitive to 
New Physics, b) from $R_b^{\rm max}=0.46$ we can
extract a robust maximum possible value for $\beta$: $|\beta|_{\rm
max}=27^\circ$, respected by the two scenarios.
 
\item $\gamma$ obtained as discussed in previous sections.
 
\item $\phi_d$ from  ${\cal A}_{\rm
CP}^{\rm mix}( B_d\to J/\psi K_{\rm S})$ is used as an input for
the CP asymmetries of $B_d \to \pi \pi $, but NOT to determine
$\beta$, since we assume that New Physics could be present.
Also $ \Delta M_d$ and
$\Delta M_s/\Delta M_d$ are not used as inputs, due to their sensitivity 
to New Physics.
 
\end{itemize}
 
Using these inputs we obtain two possible determinations for
$\alpha$, $\beta$ and $\gamma$, corresponding to the two possible 
values of $\phi_d$.

\subsection{Scenario A: Compatible with SM }
\label{sec:3}

This scenario corresponds to the first solution  
$\phi_d\!=\!47^\circ$, which implies the range for $\gamma$ given in 
Eq.~(\ref{range1}). Together with $R_b$ we obtain the black region shown 
in 
Fig 1. It implies the following prediction for the CKM angles:
$$78^\circ\leq\alpha\leq 136^\circ \quad
13^\circ\leq\beta\leq 27^\circ \quad
32^\circ\leq\gamma\leq75^\circ$$
and the error associated with $\xi\in[0.8,1.2]$ is $\Delta
\alpha=\pm4^\circ$, $\Delta
\beta=\pm1^\circ$ and $\Delta \gamma=\pm5^\circ$.
It is interesting to notice that this region is in good agreement with
the usual CKM fits~\cite{ckm}.
To illustrate it  we have shown in Fig. 1 also
the
prediction from the SM interpretation of
different
observables: $\Delta M_d$, $\Delta M_s/\Delta M_d$, $\epsilon_K$ and
 $\phi_d^{SM}=2 \beta$.

\subsection{Scenario B: New Physics}
\label{sec:4}
The second solution: $\phi_d=133^\circ$ {\it cannot } be explained in the
SM context and requires New Physics contributing to the
mixing\cite{FIM,GNW}. Models with New sources of Flavour mixing can 
account for this second solution with only two very  
general requirements~\cite{FIM}: a) The effective scale of New Physics is 
larger 
than the electroweak scale and b) the adimensional effective 
coupling ruling $\Delta B=2$ processes can always be expressed as 
the square of two $\Delta B=1$ effective couplings. Supersymmetry provides 
a perfect example, in particular, through the contribution of gluino 
mediated box diagrams with a mass insertion $\delta_{bL \; dL}^D$\cite{FIM}. 

In this case, $\gamma$
lies in the second quadrant Eq.~(\ref{range2}) and
$\beta$ is indeed smaller than in the
previous scenario. The result is still consistent with the $\epsilon_K$ 
hyperbola.
$\Delta M_{d,s}$ are not shown here, since they would be affected by New 
Physics. The
black region obtained (see Fig.2) corresponds to the following prediction
for the CKM angles:
$$22^\circ\leq\alpha\leq60^\circ \quad 8^\circ\leq\beta\leq22^\circ \quad
105^\circ\leq\gamma\leq148^\circ$$
with same errors associated to $\xi$ as in Scenario A.
It is interesting to remark that this second solution has also  
interesting implications for certain rare decays 
 like $K^+\to\pi^+\nu{\bar
\nu}$\cite{FIM,rising}. Using this second solution we find a better 
agreement with experiment than with the SM
solution. Concerning $B_{d} \to \mu^+ \mu^-$, we find also sizeable 
differences depending on the scenario used.

{\it Acknowledgements}
J.M acknowdledges financial support from FPA2002-00748.

\begin{figure}[t]
\resizebox{0.48\textwidth}{!}{%
  \includegraphics{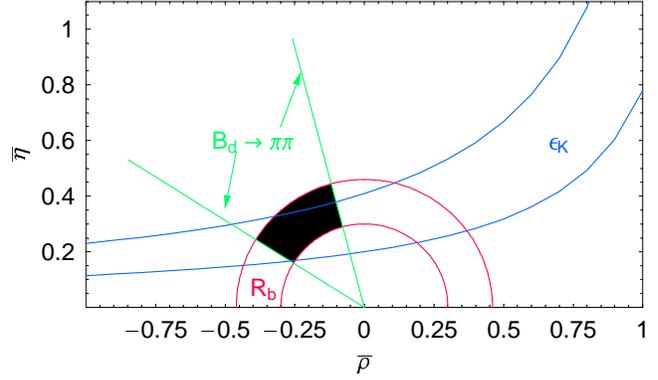}
}
\caption{$\phi_d=133^\circ$. SCENARIO B}
\label{fig:1}       
\end{figure}

%
%

\end{document}